\documentstyle[titlepage,12pt]{article}
\begin{document}
\begin{titlepage}

\rightline{}
\rightline{}
\rightline{}

\vspace*{2cm}

\addtocounter{footnote}{1}

\begin{center}

{\large \bf 
Lattice perturbation theory in the overlap formulation  \\ 

\vspace{0.5cm}

for the Yukawa and gauge interactions

}

\vspace{0.5cm}

\vspace{2cm}
{\sc Atsushi Yamada}\footnote{On leave of absence from 
Department of Physics, University of Tokyo, Tokyo, 113 Japan. }  
\\
\vspace{1cm}
{\it The Abdus Salam International Center for Theoretical Physics, \\
\vspace{0.5cm}
  Strada Costiera 11, Trieste, Italy  }
\vspace{2cm}

{\bf ABSTRACT}
\end{center}

Lattice perturbation theory is discussed in the overlap formulation 
for the Yukawa and gauge interactions. 
One and two point functions are studied for fermion, scalar and gauge 
fields, taking the Standard Model 
as an example. 
The formulae for the self-energies are given from which their divergent 
and finite parts can be computed at the one loop level. 

\end{titlepage}

\baselineskip = 0.7cm


%
%


%
%

\newcommand{\plb}[3]{Phys. Lett. {\bf B#1}, (#2), #3} 
\newcommand{\prl}[3]{Phys. Rev. Lett. {\bf #1}, (#2), #3}
\newcommand{\prd}[3]{Phys. Rev. {\bf D#1}, (#2), #3}
\newcommand{\npb}[3]{Nucl. Phys. {\bf B#1}, (#2), #3}
\newcommand{\npbps}[3]{Nucl. Phys. {\bf B}(Proc. Suppl.) {\bf #1}, (#2), #3}
\newcommand{\prog}[3]{Prog. Theor. Phys. {\bf #1}, (#2), #3}
\newcommand{\zeitc}[3]{Z. Phys. {\bf C#1}, (#2), #3}
\newcommand{\mpl}[3]{Mod. Phys. Lett. {\bf A#1}, (#2), #3} 
\newcommand{\ijmp}[3]{Int. J. Mod. Phys. Lett. {\bf A#1}, (#2), #3}

\newcommand{\be}{\begin{eqnarray}}
\newcommand{\ee}{\end{eqnarray}}

\newcommand{\gsimeq}
{\hbox{ \raise3pt\hbox to 0pt{$>$}\raise-3pt\hbox{$\sim$} }}
\newcommand{\lsimeq}
{\hbox{ \raise3pt\hbox to 0pt{$<$}\raise-3pt\hbox{$\sim$} }}

\newcommand{\delmu}{\partial_\mu}
\newcommand{\delnu}{\partial_\nu}
\newcommand{\delmunu}{\delta_{\mu\nu}}
\newcommand{\muhat}{{\hat \mu}}
\newcommand{\ran}{\rangle}
\newcommand{\lan}{\langle}
\newcommand{\tpidelta}[1]{(2\pi)^4 \delta_P(#1)}

\newcommand{\gfive}{\gamma_5}
\newcommand{\gfivetc}{\gamma_5 T_c}  
\newcommand{\fnot}[1]{\!\!\not\!#1}

\newcommand{\ptilde}{{\widetilde p}}
\newcommand{\phat}{\widehat{p}}
\newcommand{\pbar}{\bar{p}}
\newcommand{\ptildebar}{\widetilde{\bar{p}}}
\newcommand{\mf}{m_f}
\newcommand{\mfs}{m^2_f}
\newcommand{\me}{m_e}
\newcommand{\mes}{m^2_e}
\newcommand{\notp}{\!\!\not\!p}

\newcommand{\cuv}{\frac{1}{16\pi^2}\log(a^2 \mu^2)}

\newcommand{\svp}{\vert + \rangle }
\newcommand{\svm}{\vert-\rangle }
\newcommand{\svpm}{\vert \pm \rangle }
\newcommand{\svpd}{\langle +\vert}
\newcommand{\svpmd}{\langle \pm\vert}
\newcommand{\svmd}{\langle -\vert}
\newcommand{\svap}{\vert A+\rangle}
\newcommand{\svapm}{\vert A\pm\rangle}
\newcommand{\svam}{\vert A-\rangle}
\newcommand{\svapd}{\langle A+\vert}
\newcommand{\svapmd}{\langle A\pm\vert}
\newcommand{\svamd}{\langle A-\vert}

\newcommand{\psibar}{\bar{\psi}}

\newcommand{\upp}{u_{+}(p,f,\sigma_f)} 
\newcommand{\ump}{u_{-}(p,f,\sigma_f)} 
\newcommand{\vpp}{v_{+}(p,f,\sigma_f)}
\newcommand{\vmp}{v_{-}(p,f,\sigma_f)}
\newcommand{\umbp}{\bar{u}_{-}(p,f,\sigma_f)}
\newcommand{\vpbp}{\bar{v}_{+}(p,f,\sigma_f)}
\newcommand{\vmbp}{\bar{v}_{-}(p,f,\sigma_f)}

\newcommand{\upmp}{u_{\pm}(p,f,\sigma_f)}
\newcommand{\vpmp}{v_{\pm}(p,f,\sigma_f)}
\newcommand{\upmbp}{\bar{u}_{\pm}(p,f,\sigma_f)}
\newcommand{\vpmbp}{\bar{v}_{\pm}(p,f,\sigma_f)}

\newcommand{\cbep}{\cos{\beta(p,f)}}
\newcommand{\cbeq}{\cos{\beta(q,h)}}
\newcommand{\cbek}{\cos{\beta(k,g)}}

\newcommand{\splusp}{S_{+}(p,f)}
\newcommand{\smp}{S_{-}(p,f)}
\newcommand{\splusq}{S_{+}(q,h)}
\newcommand{\smq}{S_{-}(q,h)}
\newcommand{\splusk}{S_{+}(k,g)}
\newcommand{\smk}{S_{-}(k,g)}
\newcommand{\tplusp}{T_{+}(p,f)}
\newcommand{\tmp}{T_{-}(p,f)}
\newcommand{\tplusq}{T_{+}(q,h)}
\newcommand{\tmq}{T_{-}(q,h)}
\newcommand{\tplusk}{T_{+}(k,g)}
\newcommand{\tmk}{T_{-}(k,g)}
\newcommand{\pplusp}{P_{+}(p,f)}
\newcommand{\pmp}{P_{-}(p,f)}
\newcommand{\pplusq}{P_{+}(q,h)}
\newcommand{\pmq}{P_{-}(q,h)}
\newcommand{\pplusk}{P_{+}(k,g)}
\newcommand{\pmk}{P_{-}(k,g)}
\newcommand{\nplusp}{N_{+}(p,f)}
\newcommand{\nmp}{N_{-}(p,f)}
\newcommand{\nplusq}{N_{+}(q,h)}
\newcommand{\nmq}{N_{-}(q,h)}
\newcommand{\nplusk}{N_{+}(k,g)}
\newcommand{\nmk}{N_{-}(k,g)}
\newcommand{\tpluspe}{T_{+}(p,e)}
\newcommand{\tmpe}{T_{-}(p,e)}
\newcommand{\tplusqe}{T_{+}(q,e)}
\newcommand{\tmqe}{T_{-}(q,e)}
\newcommand{\tpluspse}{T_{+}(p+s,e)}
\newcommand{\tmpse}{T_{-}(p+s,e)}
\newcommand{\ttpluspe}{\widetilde{T}_{+}(\bar{p},e)}
\newcommand{\stpluspe}{\widetilde{S}_{+}(\bar{p},e)}
\newcommand{\stmpe}{\widetilde{S}_{m}(\bar{p},e)}
\newcommand{\ptpluspe}{\widetilde{P}_{+}(\bar{p},e)}
\newcommand{\ntpluspe}{\widetilde{N}_{+}(\bar{p},e)}

\newcommand{\wppq}{\omega_{+}(p,f)+ \omega_{+}(q,h)}
\newcommand{\wmpq}{\omega_{-}(p,f)+ \omega_{-}(q,h)}
\newcommand{\fwppq}{\Bigr\{ \frac{1}{\wppq} \Bigl\} }
\newcommand{\fwmpq}{\Bigr\{ \frac{1}{\wmpq} \Bigl\} }
\newcommand{\wppk}{\omega_{+}(p,f)+ \omega_{+}(k,g)}
\newcommand{\wmpk}{\omega_{-}(p,f)+ \omega_{-}(k,g)}
\newcommand{\fwppk}{\Bigr\{\frac{1}{\wppk}\Bigl\}}
\newcommand{\fwmpk}{\Bigr\{\frac{1}{\wmpk}\Bigl\}}
\newcommand{\wpkq}{\omega_{+}(k,g)+ \omega_{+}(q,h)}
\newcommand{\wmkq}{\omega_{-}(k,g)+ \omega_{-}(q,h)}
\newcommand{\fwpkq}{\Bigr\{\frac{1}{\wpkq}\Bigl\}}
\newcommand{\fwmkq}{\Bigr\{\frac{1}{\wmkq}\Bigl\}}
\newcommand{\wpp}{\omega_{+}(p,f)}   
\newcommand{\wmp}{\omega_{-}(p,f)}
\newcommand{\wpk}{\omega_{+}(k,g)}
\newcommand{\wmk}{\omega_{-}(k,g)}
\newcommand{\wpq}{\omega_{+}(q,h)}
\newcommand{\wmq}{\omega_{-}(q,h)}
\newcommand{\wpmp}{\omega_{\pm}(p,f)}   
\newcommand{\omp}{\omega_+}
\newcommand{\omm}{\omega_-}
\newcommand{\ompm}{\omega_\pm}
\newcommand{\tomp}{\tilde{\omega}_+}
\newcommand{\tomm}{\tilde{\omega}_-}
\newcommand{\tompm}{\tilde{\omega}_\pm}
\newcommand{\wppe}{\omega_{+}(p,e)}
\newcommand{\wpqe}{\omega_{+}(q,e)}
\newcommand{\wppse}{\omega_{+}(p+s,e)}
\newcommand{\twppe}{\widetilde{\omega}_{+}(\bar{p},e)}
\newcommand{\wppke}{\omega_{+}(p,e)+ \omega_{+}(k,e)}
\newcommand{\fwppke}{\Bigr\{\frac{1}{\wppke}\Bigl\}}
\newcommand{\fracom}{\frac{1}{\tilde{\omega}^2}}
\newcommand{\fracp}{\frac{1}{\tilde{\bar{p}}^2}}
\newcommand{\xp}{X_+} 
\newcommand{\xm}{X_-} 
\newcommand{\xpm}{X_\pm} 
\newcommand{\txp}{\tilde{X}_+} 
\newcommand{\txm}{\tilde{X}_-} 
\newcommand{\txpm}{\tilde{X}_\pm} 

\newcommand{\gp}{G_+}
\newcommand{\gm}{G_-}
\newcommand{\nopq}{{\bf:}\Omega(p,q){\bf:}}
\newcommand{\gampq}{\Gamma(p,q)}
\newcommand{\gamqp}{\Gamma(q,p)}
\newcommand{\gampk}{\Gamma(p,k)}
\newcommand{\gamkq}{\Gamma(k,q)}
\newcommand{\sump}{\sum_{f} \int_{p}}
\newcommand{\sumq}{\sum_{h} \int_{q}}
\newcommand{\sumk}{\sum_{g} \int_{k}}
\newcommand{\sumpq}{\sum_{f,h} \int_{p,q}}
\newcommand{\sigpq}{\Sigma(p,q)}

\newcommand{\phid}{\Phi^\dagger}
\newcommand{\pid}{\pi^\dagger}
\newcommand{\lamphi}{\lambda_\Phi}
\newcommand{\zmun}{Z_\mu(n)} 
\newcommand{\amun}{A_\mu(n)} 
\newcommand{\wpmun}{W^+_\mu(n)} 
\newcommand{\wmmun}{W^-_\mu(n)} 
\newcommand{\znun}{Z_\nu(n)} 
\newcommand{\anun}{A_\nu(n)} 
\newcommand{\wpnun}{W^+_\nu(n)} 
\newcommand{\wmnun}{W^-_\nu(n)}
\newcommand{\bmun}{B_\mu(n)} 
\newcommand{\wtmun}{W^3_\mu(n)} 
\newcommand{\zmu}{Z_\mu} 
\newcommand{\amu}{A_\mu} 
\newcommand{\wpmu}{W^+_\mu} 
\newcommand{\wmmu}{W^-_\mu} 
\newcommand{\znu}{Z_\nu} 
\newcommand{\anu}{A_\nu} 
\newcommand{\wpnu}{W^+_\nu} 
\newcommand{\wmnu}{W^-_\nu}
\newcommand{\sw}{\sin\theta_W }
\newcommand{\cw}{\cos\theta_W }
\newcommand{\mw}{M_W}
\newcommand{\mz}{M_Z}
\newcommand{\mh}{M_H}
\newcommand{\mws}{M^2_W}
\newcommand{\mzs}{M^2_Z}
\newcommand{\mhs}{M^2_H}
\newcommand{\alphaz}{\alpha_Z}
\newcommand{\alphaw}{\alpha_W}
\newcommand{\alphaa}{\alpha_A}
\newcommand{\boxhat}{\hat{ \Box}}
\newcommand{\dhat}{\hat{D}}
\newcommand{\dhatmunu}[2]{\dhat_{\mu\nu}(#1,#2)}

\renewcommand{\theequation}{\thesection.\arabic{equation}}
\setcounter{equation}{0}


\setcounter{equation}{0}
\section{Introduction}

\indent

The overlap formulation \cite{over} and Shamir type fermion \cite{shamir,neu} 
are new prescriptions for regularizing a chiral fermion on a lattice. 
Recent analytical and numerical checks on the fermion 
chirality in these formulations 
\cite{at}-\cite{atsushi} suggest 
that vector-like theories will be regularized preserving chiral symmetry, 
though there remain 
subtle points still to be clarified on the gauge invariance 
(and accordingly the renormalizability) for a regularization of a general 
chiral gauge theory. The existence of chiral symmetry allows a clearer 
study of the 
phenomenon of chiral symmetry breaking in a vector like gauge theory and 
the effects of the strong Yukawa couplings in Higgs models, and provides a 
possibility to define a chiral gauge theory in a non-perturbative way.

In this paper we discuss lattice perturbation theory for the 
Yukawa and gauge interactions in the overlap formulation. 
To be realistic, we take the Standard Model as an example, in which 
Yukawa interactions, chiral and vector like gauge interactions are 
involved, and we analyze the 
one and two point functions for fermions, scalars and gauge bosons. 
We also give the formulae for the 
self-energies at the one loop level, from which both the 
divergent and finite parts can be computed. 
Lattice perturbation theory for the Yukawa interactions 
has not been discussed before in this formulation. For the gauge interactions, 
our study supplements the existing analyses 
\cite{over} \cite{atsushi}-\cite{n}, as will be seen later. 

The rest of the paper is organized as follows. 
In Sec. 2, we define our model on a lattice, following the notation of 
Refs. \cite{atsushi,anomaly}, and discuss the 
lattice perturbation theory of the Yukawa interactions, as well as 
that of the gauge interactions in the overlap formulation. 
Both chiral and Dirac fermions are considered. 
In Sec. 3, we study the effects of the Yukawa couplings. 
The tadpole diagram and the self-energies of the 
Higgs boson and the fermion are studied. We confirm the relation between the 
tadpole term and the self-energy of the Nambu-Goldstone boson. 
In Sec. 4, we discuss the effects of the gauge interactions and compute  
the vacuum polarization and fermion self-energy. 
We demonstrate the cancellation of the quadratic divergence in the 
vacuum polarization. In these two sections, we show that the ultraviolet 
divergences are correctly reproduced and the fermion mass is renormalized 
multiplicatively. 
The general formulae for the self-energies of the fermions and 
bosons (both scalar and gauge bosons) are given in appendix A. 
In appendix B, we briefly review the lattice perturbation of the 
$SU(2)\times U(1)$ Higgs doublets. 
 
\setcounter{equation}{0}

\section{Formalism}

\indent

In the overlap formulation, the effective action due to chiral fermions 
in the presence of gauge and scalar fields is expressed by the overlap 
of the two vacua $\svapm$ of the two Hamiltonians ${\cal H}_\pm$.     
These two Hamiltonians are different from each other only in the sign of the 
mass term, and are derived from the theory describing 4+1 dimensional Dirac 
fermions, which simulate chiral fermions in 4 dimensions. 
Consider a multiplets $\psi$ of massive Dirac fermions in 4+1 
dimensions, interacting with the Higgs doublet 
\be
\Phi(x) = \left(  
\begin{array}{c}
\phi_+(x) \\
\phi_0(x)
\end{array}
\right)
=
\left(  
\begin{array}{c}
i \pi_+(x) \\
\{ v + H(x) -i \pi_3(x)  \}/{\sqrt{2}}
\end{array}
\right).
\label{eqn:hdoub}
\ee
Here the Higgs fields do not depend on the fifth coordinate 
(which is the time in 4+1 dimensional space-time).  
The Hamiltonians are given by 
\be
& &{\cal H}_{\pm} = \int d^4 x \psi^\dagger(x) \gfive 
\Bigl[\sum_{\mu=1}^{4}  \gamma_\mu \partial_\mu   
\pm  T_c \Lambda  + H_Y(x) \Bigr]  \psi(x),
\label{eqn:hcon}
\ee
where the mass term $\Lambda$ corresponds to the height of the domain wall 
in the original domain wall fermion \cite{kaplan,chiral}, 
$T_c$ is the chirality matrix which 
determines the chirality of each components of $\psi$, and $H_Y(x)$ 
describes the interaction of the chiral multiplet with Higgs 
scalars \cite{anomaly}. For the lepton sector of the standard model, 
\be
\psi= 
\left( 
\begin{array}{c} 
\nu_L \\
e_L \\
e_R 
\end{array}
\right),\,\,\,\,\, 
T_c = 
\left( 
\begin{array}{ccc} 
-1  &  0 &  0  \\
 0  & -1 &  0  \\
 0  &  0 &  1  
\end{array}
\right),\,\,\,\,\, 
H_Y = y_l 
\left( 
\begin{array}{ccc} 
 0          &  0       &  \phi_+  \\
 0          &  0       &  \phi_0  \\
 \phi^*_+   &  \phi^*_0  &    0  
\end{array}
\right), 
\label{eqn:sml}
\ee
and for the quark sector of the standard model,  
\be
& &\psi= 
\left( 
\begin{array}{c} 
u_L \\
d_L \\
u_R \\
d_R
\end{array}
\right),\,\,\,\,\, 
T_c = 
\left( 
\begin{array}{cccc} 
-1  &  0 &  0  &  0 \\
 0  & -1 &  0  &  0 \\
 0  &  0 &  1  &  0 \\
 0  &  0 &  0  &  1
\end{array}
\right),\,\,\,\,\, 
\nonumber \\
& & H_Y = 
y_u
\left( 
\begin{array}{cccc} 
    0    &      0   &  \phi^*_0  &  0 \\
    0    &      0   & -\phi^*_+  &  0 \\
 \phi_0  &  -\phi_+ &    0       &  0 \\
    0    &      0   &    0       &  0
\end{array}
\right) 
+
y_d
\left( 
\begin{array}{cccc} 
      0      &      0   &   0  &  \phi_+ \\
      0      &      0   &   0  &  \phi_0 \\
      0      &      0   &   0  &    0    \\
    \phi^*_+ & \phi^*_0 &   0  &    0
\end{array}
\right),\,\,\,\,\, 
\label{eqn:smq}
\ee
where $y_l$, $y_u$ and $y_d$ are the Yukawa couplings.
Since the Higgs sector connects the left-handed fermions to the right-handed  
fermions and vise versa, the chirality matrix $T_c$ anti-commutes with 
$H_Y$; $ \{T_c,H_Y\}=0$. 
From the Hamiltonians (\ref{eqn:hcon}), 
the gauge invariant Hamiltonians on the lattice are obtained 
as
\be
{\cal H}_{\pm} &=&  
a^4\sum_{n} \psi^\dagger_n  \gfive
\Bigl[
\frac{1}{2a} \sum_{\mu}   [ 
\gamma_\mu 
\Bigl\{ 
W_{n,n+\muhat}\psi_{n+\muhat}
-W_{n,n-\muhat}\psi_{n-\muhat}
\Bigr\}
\nonumber \\
& &+ T_c\Bigl\{ \lambda/a \psi_{n}   \pm \frac{r}{2a}  \sum_{\mu}  
\Bigl( 
 W_{n,n+\muhat}\psi_{n+\muhat}
+W_{n,n-\muhat}\psi_{n-\muhat}
-2 \psi_{n}
\Bigr)
\Bigr\}+
H_Y \psi_{n}
\Bigr],
\label{eqn:hlat}
\ee
by discretising Eq. (\ref{eqn:hcon}), 
adding the Wilson term and inserting the link variable.
The link variable $W_{n,n+\muhat}$ is defined as
$W_{n,n+\muhat}=U_{n,n+\muhat}V_{n,n+\muhat}$, where 
$U_{n,n+\muhat}$ and $V_{n,n+\muhat}$ are the link variables of the 
$SU(2)$ and $U(1)$ gauge interactions, respectively. They commute with each 
other and satisfy the 
relations $U_{n,n+\muhat}=U^\dagger_{n+\muhat,n}$ and  
$V_{n,n+\muhat}=V^\dagger_{n+\muhat,n}$. 
 
To quantize the system, we set the commutation relations,
\be  
\{\psi_{\alpha m}, \psi^\dagger_{\beta n} \} = 
\frac{1}{a^4} \delta_{\alpha\beta} \delta_{mn},
\,\,\,\,\, 
\{\psi_{\alpha m}, \psi_{\beta n} \} =
\{\psi^\dagger_{\alpha m}, \psi^\dagger_{\beta n} \} =0, 
\label{eqn:etc}
\ee
which are the equal time commutation relations for Dirac fermions 
in 4+1 dimensions. 
 
To develop the perturbation theory, 
we divide the Hamiltonian (\ref{eqn:hlat}) into the free part and 
the interactions, and go into the momentum space. 
$H_Y$ is divided into the mass matrix $M$ and 
the interaction part as $H_Y= M + Y_n$, $M=\lan H_Y \ran$,  
where $\lan H_Y \ran$ contains only the vacuum expectation value of the 
doublet (\ref{eqn:hdoub}). 
The link variables are expanded in terms of the gauge couplings $g$ 
and $g'$ using the identifications, 
\be
U_{n,n+\muhat}= e^{iga\sum_{i}T^iW^i_\mu(n)},\,\,\,\,\,  
V_{n,n+\muhat}= e^{ig'a Y B_\mu(n)}, 
\label{eqn:exlink}
\ee
where $T^i$ and $Y$ are the generators of the $SU(2)$ and 
$U(1)$ gauge group corresponding to the multiplet $\psi$. 
For the lepton sector, $T^i$ and $Y$ are three by three matrices given by  
\be
T^i = 
\left( 
\begin{array}{cc}
\frac{1}{2}\sigma_i & 0 \\
       0            & 0                   
\end{array} 
\right),\,\,\,\,\, 
Y  = 
\left( 
\begin{array}{ccc}
 Y_L & 0    &   0  \\
 0   & Y_L  &   0  \\
 0   & 0    &  Y_{l_R}              
\end{array} 
\right), 
\ee
and for the quark sector, $T^i$ and $Y$ are four by four 
matrices given by 
\be
T^i = 
\left( 
\begin{array}{ccc}
\frac{1}{2}\sigma_i & 0 & 0 \\
       0            & 0 & 0 \\                 
       0            & 0 & 0 
\end{array} 
\right),\,\,\,\,\, 
Y  = 
\left( 
\begin{array}{cccc}
 Y_Q & 0    &  0      &  0  \\
 0   & Y_Q  &  0      &  0  \\
 0   & 0    & Y_{u_R} &  0  \\ 
 0   & 0    &  0      & Y_{d_R}               
\end{array} 
\right), 
\ee
where $\sigma_i$ are the Pauli matrices and 
$Y_{l_L}=-1/2$, $Y_{l_R}=-1$, $Y_Q = 1/6$, $Y_{u_R}=2/3$ and 
$Y_{d_R}=-1/3$. 
Performing the following Fourier transformations,  
\be
\psi_n= \int_{p} \psi(p) e^{ipan}, 
\,\,\,
\psibar_n= \int_{q} \psibar(q) e^{-iqan}, 
\,\,\,
A^i_\mu (n) =\int_{p} A^i_\mu(p) e^{ipa(n + \hat{\mu}/2)} ,
\,\,\,
Y_n=\int_{p} Y(p) e^{ipan}, 
\nonumber \\
\label{eqn:fourier}
\ee
the Hamiltonians (\ref{eqn:hlat}) are divided into the free part and 
the interaction part as, 
\be
& &{\cal H}_{\pm}(A.\Phi) = \int_{p} \psi^\dagger (p) H_{\pm}(p) \psi(p) +
{\cal V}(A,\Phi), 
\label{eqn:H}
\\
& &H_{\pm}(p)=\gfive\Bigl[\sum_{\mu} i {\tilde p}_\mu \gamma_\mu + T_c
X_{\pm}(p) + M     \Bigr] , \,\,\,\,\,\,
X_{\pm}(p)= \pm\frac{\lambda}{a} + \frac{ar}{2} \hat{p}^2,
\label{eqn:h}
\ee
where $\tilde{p}_\mu=(1/a)\sin (p_\mu a)$, $\hat{p}_\mu=(2/a)\sin(p_\mu a/2)$ 
and the momentum integral is over the Brillouin zone $[-\pi/a,\pi/a]$.

First we consider the free part and then take the interactions into account as perturbations. To quantize the free part, 
we expand the operator $\psi (p)$ in terms of the creation and 
annihilation operators as 
\be
\psi(p)=  \sum_{f,\sigma-f} \Bigl[ \upmp b_{\pm}(p,f,\sigma_f)
+ \vpmp d^\dagger_{\pm}(p,f,\sigma_f)  \Bigr],
\label{eqn:psi}
\ee     
where $\upmp$ and $\vpmp$ are the eigenspinors of the one-particle 
free Hamiltonians $H_{\pm}(p)$ and satisfy the eigenequations  
\be
& &H_{\pm}(p) u_{\pm}(p,f,\sigma_f) = 
\omega_\pm (p,f) u_{\pm}(p,f,\sigma_f),
\,\,\, 
H_{\pm}(p) v_{\pm}(p,f,\sigma_f) = -\omega_\pm (p,f) v_{\pm}(p,f,\sigma_f), 
\nonumber \\
& &\,\,\,\,\,\,\,\,\,\,\,\,\,\,\,\,\,\,
\omega_\pm (p,f) =\sqrt{{\tilde p}^2+ m^2_f + X^2_{\pm}(p) }. 
\label{eqn:spinor}
\ee
Here $\mf^2$ are the eigenvalues of the $M^2$ and 
we denote the flavor indices of the spinors 
as $f$ and their spin indices as $\sigma_f$. 
For example, $f=\nu_L$, $e_L$ and $e_R$ for leptons.
In Eq. (\ref{eqn:spinor}), the relations $m_{e_L}=m_{e_R}$ and 
$\omega_\pm (p,e_L)=\omega_\pm (p,e_R)$ hold. 
(Later we will introduce the propagator for the Dirac fermion with the 
index $e_V$, which is the sum of the propagators of $e_L$ and $e_R$.) 
The explicit expressions of the eigenspinors are 
\be
\upmp =
\frac{\{\omega_{\pm}(p,f) + H_{\pm}(p)\}}
{\sqrt{2\omega_{\pm}(\omega_{\pm} + X_{\pm})}} \chi(f,\sigma_f),
\nonumber \\
\vpmp =
\frac{\{\omega_{\pm}(p,f) - H_{\pm}(p)\} }
{\sqrt{2\omega_{\pm}(\omega_{\pm} - X_{\pm})}} \chi(f,\sigma_f),
\label{eqn:uv}
\ee
where the spinors $\chi(f,\sigma_f)$ satisfy 
$\gfivetc \chi(f,\sigma_f)=\chi(f,\sigma_f)$ and 
$M^2 \chi(f,\sigma_f)=m^2_f \chi(f,\sigma_f)$. 
The orthonormalioty conditions for these spinors are 
\be
u_{\pm}(p,f,\sigma_f)^\dagger  u_{\pm}(p,g,\sigma_g)=  
v_{\pm}(p,f,\sigma_f)^\dagger  v_{\pm}(p,g,\sigma_g)= 
\delta_{fg}\delta_{\sigma_f \sigma_g },
\,\,\,\,\,
u_{\pm}(p,f,\sigma_f)^\dagger  v_{\pm}(p,g,\sigma_g)= 0.
\nonumber \\
\label{eqn:ortho}
\ee
The creation annihilation operators $(b_+,d_+) $ satisfy the 
commutation relations
\be
& &\{b_{+}(p,f,\sigma_f),b^\dagger_{+}(q,g,\sigma_g)\}= 
\{d_{+}(p,f,\sigma_f),d^\dagger_{+}(q,g,\sigma_g)\}=
(2\pi)^4\delta_{f,g}\delta_{\sigma_f\sigma_g}\delta^4_P(p-q),
\nonumber \\
& &\{b_{+}(p,f,\sigma_f),d^\dagger_{+}(q,g,\sigma_g)\}=
\{b_{+}(p,f,\sigma_f),d_{+}(q,g,\sigma_g)\}=0,
\label{eqn:cm}
\ee
where $\delta_P(p-q)$ is the periodic $\delta$-function on the lattice. 
(The same relations hold also for $(b_-,d_-) $). 
The two spinor basis $(u_+,v_+)$ and $(u_-,v_-)$ are related by 
the transformations, 
\be
& &\ump= \cbep \upp -\sin\beta(p,f) \vpp,
\nonumber \\
& &\vmp= \sin\beta(p,f) \upp + \cbep \vpp,
\label{eqn:bog1}
\ee
where $\cbep= u^\dagger_{+}(p,f,\sigma_f) \ump$ is given by, 
\be
\cbep =\frac{1}{\sqrt{2\omega_+ 2\omega_-}}
[\sqrt{(\omega_+ + X_+)(\omega_- +  X_-) }  +
\sqrt{(\omega_+ - X_+)(\omega_- -  X_-) }
].
\label{eqn:cb}
\ee
Eqs. (\ref{eqn:bog1}) lead to the Bogoluibov transformation 
\be
& &b_{-}(p,f,\sigma_f)= \cbep b_{+}(p,f,\sigma_f) -\sin\beta(p,f)
d^\dagger_{+}(p,f,\sigma_f),
\nonumber \\
& &d^\dagger_{-}(p,f,\sigma_f)= 
\sin\beta(p,f) b_{+}(p,f,\sigma_f)+ \cbep
d^\dagger_{+}(p,f,\sigma_f).
\label{eqn:bog}
\ee
between the two basis $(b_+,d_+)$ and $(b_-,d_-) $.

The free propagator is defined by the vacuum expectation value of the 
operator 
$\Omega(p,q) =  \{\psi(p) \psibar(q) - \psibar(q)\psi(p)\}/2$ as, 
\be
\svpd \Omega(p,q) \svm &=&  (2\pi)^4 \delta^{4}_P(p-q)
\sum_{f}\frac{1}{2}\Bigl[ S_{+}(p,f) -S_{-}(p,f)  \Bigr],
\label{eqn:tree}
\nonumber \\
S_+ (p,f) &=&
\frac{1}{\cbep} \sum_{\sigma_f} \upp\umbp,     
\label{eqn:splus}
\nonumber \\
S_-(p,f)  &=& 
\frac{1}{\cbep} \sum_{\sigma_f} \vmp\vpbp.     
\label{eqn:sminus}
\ee 
The pole of the propagator is the zero of $\cbep$, which occurs if both 
the two relations $\omp=\xp$ and $\omm=-\xm$ are satisfied. 
This condition is fulfilled only at the center of the Brillouin zone.  
Since $\cos\beta(p,f_L)= \cos\beta(p,f_R)$, it is convenient to 
introduce the Dirac type propagator 
$S_\pm (p,f_V)= S_\pm (p,f_L) + S_\pm (p,f_R)$. 

The spinors $u_\pm$ and $v_\pm$ have the following continuum limits 
at the center of the Brillouin zone: 
\be
& &\upp = 
[1+\frac{a}{2\lambda} \{ i\notp + M \} + {\cal O}(a^2)]\chi(f,\sigma_f), 
\nonumber \\
& &\ump= 
\frac{1}{\sqrt{p^2+m^2_f}}
[ \{ i\notp + M \}  + \frac{a}{2\lambda}(p^2+\mf^2) +{\cal O}(a^2)]
\chi(f,\sigma_f),
\nonumber \\
& &\vmp= 
[1+\frac{a}{2\lambda} \{ i\notp + M \} + {\cal O}(a^2)]\chi(f,\sigma_f), 
\nonumber \\
& &\vpp= 
\frac{1}{\sqrt{p^2 +\mf^2 }} 
[- \{ i\notp + M \}  + \frac{a}{2\lambda} (p^2+\mf^2) +{\cal O}(a^2)]
\chi(f,\sigma_f),
\label{eqn:uvl}
\ee
and the continuum limit of $\cbep$ is 
\be
\cbep={\frac{a}{\lambda} \sqrt{p^2+\mf^2}} +{\cal O}(a^3). 
\ee
For the lepton sector, the continuum limits of the propagators are 
\be
& &S_{\pm}(p,\nu_L) = 
\frac{\lambda}{a}\frac{1}{p^2}
\left(
\begin{array}{ccc} 
\mp P_L i\notp & 0 & 0 \\
     0     & 0 & 0 \\
     0     & 0 & 0 
\end{array}
\right),
\nonumber \\
& &S_{\pm}(p,e_V) = 
\frac{\lambda}{a}\frac{1}{p^2+m^2_e}
\left(
\begin{array}{ccc} 
     0     &       0        &           0       \\
     0     & \mp P_L i\notp &    \pm P_L m_e    \\
     0     & \pm P_R m_e    &    \mp P_R i\notp  
\end{array}
\right),
\ee
where $P_{L(R)} =(1-(+)\gfive)/2 $.   

Next consider the interaction parts. 
The Yukawa interaction is given by 
\be
V_Y = a^4 \sum_{n} \psibar_n Y_n \psi_n 
=\int_{p,q} \psibar(p) Y(p-q) \psi(q). 
\label{eqn:yukawa}
\ee
For the lepton sector, 
\be
Y(p)=y_l
\left(
\begin{array}{ccc}
0&0& i\pi_+(p)  \\
0&0& \{H(p)-i\pi_3(p)\}/\sqrt{2} \\
-i\pi^*_+(p)& \{H(p)+i\pi_3(p)\}/\sqrt{2} &0 
\end{array}
\right). 
\ee

The gauge interactions are obtained from the expansion of the link variable 
\be
W_{n,n+\muhat} &=& 1 +ia 
\Bigl[
eQ \amun + g_z (T^3- s^2_w Q) \zmun
+\frac{g}{\sqrt{2}} \Bigl\{ T_+ \wpmun +   T_- \wmmun \Bigr\}
\Bigr] 
\nonumber \\
&+&\frac{1}{2!} (ia)^2 
\Bigl[
eQ \amun + g_z (T^3- s^2_w Q) \zmun
+\frac{g}{\sqrt{2}} \Bigl\{ T_+ \wpmun +   T_- \wmmun \Bigr\}
\Bigr]^2 + \cdots 
\nonumber \\
\ee
as
\be 
H_G  &=&i \int_{p,q} \sum_{\mu} \psibar(p)  
V_{1\mu}(p+q) G_\mu(p-q) \psi(q) 
\nonumber\\
&+&
\frac{1}{2!}a\int_{p,s,q} 
\psibar(p)  \sum_{\mu\nu} \delmunu
V_{2\mu}(p+q) G_\mu(s) G_\mu(p-s-q) \psi(q) +\cdots 
\ee
where 
$V_{1\mu}(p) = \gamma_{\mu} \cos(p_\mu a/2) -ir T_c \sin(p_\mu a/2)$,  
$V_{2\mu}(p)=  T_c r\cos(p_\mu a/2) -i\gamma_\mu \sin(p_\mu a/2)$
and 
\be
G_\mu(p) = eQ \amu(p) + g_z (T^3- s^2_w Q) \zmu(p)
+\frac{g}{\sqrt{2}} \Bigl\{ T_+ \wpmu(p) +   T_- \wmmu(p) \Bigr\}.
\ee
Here $Q = T^3 + Y$ and $T_\pm = T^1 \pm iT^2$. 

Now we discuss the perturbation theory. 
The Dirac vacua $\svpm$ for the free Hamiltonians 
$ {\cal H}_{\pm}(A=0)$ are defined as, $b_{+},d_{+}\svp =0$ and
$b_{-},d_{-}\svm =0$ and their energy eigenvalues
are denoted as $E_\pm(0)$, 
$ {\cal H}_{\pm}(A=0) \svpm = E_\pm(0) \svpm$. 
Then for the Dirac vacua $\svapm$, the eigenvalue equations, 
\be
{\cal H}_{\pm}(A) \svapm = E_{\pm}(A) \svapm, 
\label{eqn:eigena}
\ee
are solved in the form of the integral equation using the 
Dirac vacua $\svpm$ following the standard time independent 
perturbation theory. The results are,   
\be 
\svapm &=& \alpha_\pm(A)\Bigl[1-G_\pm 
({\cal V}-\Delta E_\pm)\Bigr]^{-1} \svpm,
\label{eqn:al}
\ee
where, $\Delta E_\pm = E_{\pm}(A)- E_{\pm}(0)= 
\svpmd {\cal V} \svapm / \svpmd A\pm \ran$ and  
\be
G_\pm = \sum_{n} {}' \vert n \pm \rangle \frac{1}{E_\pm(0)-E_\pm(n)} 
\langle n \pm \vert =  
\frac{1-\svpm \svpmd}{E_\pm(0)-H_\pm(0)} ,
\label{eqn:gpm}
\ee
In Eq. (\ref{eqn:gpm}), 
the sum $ \sum_{n}'$ is over all the excited states 
$\vert n \pm \rangle$ of ${\cal H}_{\pm}(0) $ and 
$ E_\pm(n)$ denote their energy eigenvalues. 
The normalization factors $\alpha_\pm(A)$ are determined, up to phases 
\cite{phase}, 
by the normalization conditions of 
$\svapm$ as 
\be
\vert \alpha_\pm(A)   \vert^2= 1-
\svapmd \Bigl[{\cal V}-\Delta E_\pm\Bigr]G^2_\pm
\Bigl[{\cal V}-\Delta E_\pm\Bigr] \svapm. 
\label{eqn:alpha}
\ee
 
The correlation functions for the $\psi$, $\anu$ and $\Phi$ 
fields are obtained in the form of path integral. 
For example, the two point function for $\psi$ is given by 
\be 
\frac{\int {\cal D}{\cal A}
\langle A+ \vert \Omega(p,q)\vert A- \rangle e^{-S(A)} }
{ \int {\cal D}{\cal A}    \langle A+ \vert A- \rangle e^{-S(A)} },
\label{eqn:twof}
\ee
and the two point function for the gauge field is given by 
\be 
\frac{\int {\cal D}{\cal A} {\cal D} \Phi 
\amu(s)\anu(t) \langle A + \vert  A- \rangle e^{-S(A)} }
{ \int {\cal D}{\cal A}    \langle A+ \vert A- \rangle e^{-S(A)} },
\label{eqn:twog}
\ee
where $S$ is the actions for the gauge fields and Higgs fields, 
including the effects of the path integral measure, gauge fixing terms and 
Faddeev-Popov ghost terms. 
To evaluate Eqs. (\ref{eqn:twof}) and (\ref{eqn:twog}), 
$\svapm$ should be expanded in terms of $V$ using 
Eqs. (\ref{eqn:al}) and (\ref{eqn:gpm}), where $V$ is 
given by the bilinear of the fermion operator. Then the vacuum 
expectation value is computed from the commutation relations 
(\ref{eqn:etc}). Finally, the path integral of over the gauge and scalar 
fields are performed following the standard perturbation theory. 
We list the results of such evaluations for 
$\langle A+ \vert \Omega(p,q)\vert A- \rangle$ and 
$\langle A + \vert  A- \rangle$ in appendix A, from which we can 
compute the fermion and boson two point functions. 
The propagators of the gauge and Higgs fields are derived in appendix B.


\setcounter{equation}{0}

\section{Yukawa Interactions}

\indent

In this section, we study the Yukawa interactions.
First we consider the contribution of the electron to the 
tadpole of the Higgs field    
\be 
\frac{\int {\cal D}{\cal A} {\cal D} \Phi 
H(s) \langle A + \vert  A- \rangle e^{-S(A)} }
{ \int {\cal D}{\cal A}    \langle A+ \vert A- \rangle e^{-S(A)} }. 
\label{eqn:htad}
\ee
Here $\langle A + \vert  A- \rangle$ should be expanded to the first order 
in the Yukawa couplings: 
\be
\langle A + \vert  A- \rangle = \svpd V \gp  \svm + \svpd \gm V \svm,
\ee
and $V$ is given by Eq. (\ref{eqn:yukawa}) with 
\be 
Y(p-q) =\frac{y}{\sqrt{2}} H(p-q) 
\left(
\begin{array}{ccc}
 0 & 0 & 0  \\
 0 & 0 & 1  \\
 0 & 1 & 0 
\end{array}
\right). 
\label{eqn:yukawah}
\ee
Using the expressions (\ref{eqn:vgpb}) and (\ref{eqn:gmvb}) in appendix A, 
\be 
\svpd V \gp  \svm + \svpd \gm V \svm 
&=&
-\sump \frac{1}{2\wpp} Tr Y(0) \tplusp
\nonumber \\
&+&
\sump \frac{1}{2\wmp} Tr Y(0) \tmp,
\label{eqn:tadh2}
\ee
where
\be
Tr Y(0) T_\pm(p,f) 
=\frac{1}{\sqrt{2}}yH(0)
Tr
\Bigl\{
T_\pm(p,e)_{LR} + T_\pm(p,e)_{RL}
\Bigr\},
\ee
and the trace in the right-hand side 
is over the Dirac matrices (The subscripts ``$LR$'' 
etc indicate the elements of the matrix propagator $T_\pm$. 
See Eq. (\ref{eqn:spmex})).   
From the explicit expressions for the propagators in the appendix A,  
we get the tadpole term $H(0) \{t_+ + t_-\}$ with
\be 
& &t_+= \frac{1}{\sqrt{2}} y m_e \int_{p}
\frac{1}{\omp} 
\Bigl[  
\frac{1}{Z_+} 
\Bigl\{  
\omp+\xp +\omm+\xm
\Bigr\}
-\frac{1}{\omp}
\Bigr],
\nonumber\\ 
& &
t_-= \frac{1}{\sqrt{2}} y m_e \int_{p}
\frac{1}{\omm} 
\Bigl[  
\frac{1}{Z_-} 
\Bigl\{  
\omm-\xm +\omp-\xp
\Bigr\}
-\frac{1}{\omm}
\Bigr].
\ee
Rescaling the loop momentum as $p \rightarrow \pbar=ap$, 
we find 
$\ompm \rightarrow \widetilde{\omega}_\pm/a$ and 
$\xpm \rightarrow \widetilde{X}_\pm/a$, 
where 
\be 
\widetilde{\omega}_\pm(\pbar,e) = 
\sqrt{ \widetilde{\pbar}^2+ a^2m^2_e + \widetilde{X}_\pm(\pbar) },
\,\,\,\,\,\,\,
\widetilde{X}_\pm(\pbar)=
\pm \lambda + \frac{r}{2} \hat{\pbar}^2,
\ee
with $\ptildebar_\mu =\sin\pbar_\mu $ and 
$ \hat{\pbar}=2\sin(\pbar_\mu/2) $.  
So the tadpole term is in fact quadratically divergent; 
$t_\pm=\tilde{t}_\pm /a^2$. This term should be canceled by the 
tadpole counter-term in the action of the Higgs boson 
(\ref{eqn:shg}) given in appendix B: 
$e^{-S} \sim -\delta(\mu^2 +\lamphi v^2) v H(0)$, as 
\be
-\delta(\mu^2 +\lamphi v^2)v + 
\frac{1}{a^2} ( \tilde{t}_+ + \tilde{t}_- ) = 0. 
\ee

Because of the global $SU(2)$ symmetry, 
this quadratic divergence of the tadpole counter-term is related to the 
mass counter-terms of the Nambu-Goldstone bosons $\pi_\pm$ and $\pi_3$. 
For example, 
\be
\frac{1}{a^2}(\tilde{t}_+ + \tilde{t}_-)  = v \Sigma_{\pi_3\pi_3}(0). 
\label{eqn:tadself}
\ee
Here the self-energy of $\pi_3$ is computed from  
\be 
\frac{\int {\cal D}{\cal A} {\cal D} \Phi 
\pi_3(s)\pi_3(t) \langle A + \vert  A- \rangle e^{-S(A)} }
{ \int {\cal D}{\cal A}    \langle A+ \vert A- \rangle e^{-S(A)} }, 
\label{eqn:selfpi}
\ee
with (up to the one loop level)
\be 
\svapd A-\rangle &=& 
\Bigr\{ 1  -  \frac{1}{2} \svpd V \gp^2  V \svp  
- \frac{1}{2} \svmd V \gm^2  V \svm \Bigl\} \svpd -\rangle
\nonumber \\
&+&
\svpd V \gp \gm V  \svm 
+
\svpd V \gp  V \gp   \svm 
+
\svpd  \gm V \gm V   \svm.  
\label{eqn:selfpi2}
\ee
(In this case, $\svpd V\gp\svm$ and $\svpd \gm V \svm$ in 
Eq. (\ref{eqn:loopb}) are the one-particle reducible contributions 
which should be already renormalized, so we did not show them here.) 
The interaction $V$ is given by (\ref{eqn:yukawa}) with   
\be 
Y(p-q) =\frac{-i y}{\sqrt{2}} \pi_3(p-q) 
\left(
\begin{array}{ccc}
 0 &  0 & 0  \\
 0 &  0 & 1  \\
 0 & -1 & 0 
\end{array}
\right). 
\ee
As an example, we compute the term: $\svpd V \gp  V \gp   \svm$. 
For the first term of Eq. (\ref{eqn:vgpvgpb}), 
\be 
\svpd V \gp  V \gp   \svm &=& 
\frac{1}{4} y ^2 \int_{p,q} 
\Bigl\{ 
\frac{1}{\wppe +\wpqe} 
\Bigr\}^2
\pi_3(p-q) \pi_3(q-p)   
\nonumber \\
&\times &
Tr \tpluspe       
\left(
\begin{array}{cc}
0  & 1 \\
-1 & 0 
\end{array}
\right)
\tplusqe 
\left(
\begin{array}{cc}
0  & 1 \\
-1 & 0 
\end{array}
\right). 
\ee
Inserting this expression into (\ref{eqn:selfpi}) and performing the 
path integral, the bilinears of the $\pi_3$ fields should be 
replaced by its propagator given in appendix B, and we get 
$\tpidelta{s+t}D(s) \Sigma_{\pi_3\pi_3}(s) D(s) $ with
\be 
\Sigma_{\pi_3\pi_3}(s)= \frac{1}{2} y^2 
\int_{p} 
\Bigl\{ 
\frac{1}{\wppe +\wppse} 
\Bigr\}^2
Tr \tpluspe       
\left(
\begin{array}{cc}
0  & 1 \\
-1 & 0 
\end{array}
\right)
\tpluspse 
\left(
\begin{array}{cc}
0  & 1 \\
-1 & 0 
\end{array}
\right).
\nonumber \\
\ee
Rescaling the loop momentum as $p \rightarrow \pbar =ap $, we get the 
quadratic divergence, 
\be
\Sigma_{\pi_3\pi_3}(0) =-\frac{1}{a^2}y^2
\int_{p}
\frac{1}{4 \twppe^2} Tr \ttpluspe_{LL} \ttpluspe_{RR},  
\ee
where $\widetilde{T}_+$ is the propagator after the momentum rescaling. 
Similarly, the second part of Eq. (\ref{eqn:vgpvgpb}) 
yields the following quadratic divergence,
\be 
\Sigma_{\pi_3\pi_3}(0) &=&
\frac{1}{a^2}y^2\int_{p}
\frac{1}{4 \twppe^2} Tr 
\Bigl[ 
\ttpluspe_{LL} 
\Bigl\{  
  \ptpluspe_{RR} -\ntpluspe_{RR} 
\Bigr\}         
\nonumber\\
&+&
\ttpluspe_{RR} 
\Bigl\{  
  \ptpluspe_{LL} -\ntpluspe_{LL} 
\Bigr\}    
\Bigr]. 
\ee
The contributions of the other terms in Eq. (\ref{eqn:selfpi2}) 
are computed similarly, and the quadratic divergence of 
$\Sigma_{\pi_3\pi_3}(0)$ should satisfy the relation 
(\ref{eqn:tadself}). 
For simplicity, we confirm this relation in the limit 
of the Wilson parameter $r \rightarrow 0$. In this limit, 
$\tomp=\tomm$ and $\txm=-\txp$, and each term in 
Eq. (\ref{eqn:selfpi2}) leads to the following contributions to 
$\Sigma_{\pi_3\pi_3}(0)$, 
\be
\svpd V \gp  V \gp   \svm,\,\,\, \svpd  \gm V \gm V   \svm
&\rightarrow &
\Sigma_{\pi_3\pi_3}(0)= 
-\frac{1}{2a^2} y^2\int_{p} 
\Bigl\{ 
\fracom -\fracp 
\Bigr\}  
\nonumber \\
- \frac{1}{2} \svpd V \gp^2  V \svp,\,\,\,  
- \frac{1}{2} \svmd V \gm^2  V \svm 
&\rightarrow &
\Sigma_{\pi_3\pi_3}(0)= 
-\frac{1}{2 a^2}y^2 \int_{p} 
\fracom  
\nonumber \\
\svpd V \gp \gm V  \svm 
&\rightarrow &
\Sigma_{\pi_3\pi_3}(0)= 
\frac{y^2}{a^2} \int_{p} 
\fracp  
\ee
while the tadpole contribution is  
\be
(t_+ + t_- ) =
\frac{1}{a^2}
2\sqrt{2}y \me  \int_{p} 
\Bigl\{ 
\fracp -\fracom 
\Bigr\}.  
\ee
Now it is easy to see that the relation (\ref{eqn:tadself}) is satisfied. 


Next we compute the Higgs boson self-energy given by 
\be 
\frac{\int {\cal D}{\cal A} {\cal D} \Phi 
H(s)H(t) \langle A + \vert  A- \rangle e^{-S(A)} }
{ \int {\cal D}{\cal A}    \langle A+ \vert A- \rangle e^{-S(A)} }, 
\ee
with $\langle A + \vert  A- \rangle$ given by Eq. (\ref{eqn:selfpi2}). 
We compute the contribution of the first term of 
Eq. (\ref{eqn:vgpvgpb}). 
Performing the path integral, it is reduced to the form of    
$\tpidelta{s+t}D(s) \Sigma_{HH}(s) D(s) $ with 
\be 
\Sigma_{HH}(s)= -\frac{1}{2}y^2
\int_{p} 
\Bigl\{ 
\frac{1}{\wppe +\wppse} 
\Bigr\}^2
Tr \tpluspe       
\left(
\begin{array}{cc}
0  & 1 \\
1 & 0 
\end{array}
\right)
\tpluspse 
\left(
\begin{array}{cc}
0  & 1 \\
1 & 0 
\end{array}
\right).
\nonumber \\
\ee
Here $\Sigma_{HH}(s)$ contains the quadratic and logarithmic divergences, as 
well as constant terms. 
The quadratically divergent part is same to $\Sigma_{\pi_3 \pi_3}(0)$. 
The logarithmic divergence is evaluated by dividing the integration region 
into two pieces, following Ref. \cite{smit}. 
After the rescaling of the loop momentum $p \rightarrow \pbar=ap$, the 
logarithmic (infrared) divergence occurs in the limit 
$a \rightarrow 0$ near the center of the Brillouin zone. 
(The other corners of the Brillouin zone do not leads to the pole of the 
propagators.) Dividing the integration region into 
the vicinity of the center of the Brillouin zone and elsewhere,   
in the calculation of the former region, 
the propagators can be expanded in terms of $\pbar$ and $a$, leading to the 
expression, 
\be
\Sigma_{HH}(s)-\Sigma_{HH}(0) &=& 
\frac{y^2}{2a^2} \int 
\frac{\pbar^2 + a\pbar s - a^2 \mes }
{(\pbar^2+a^2 \mes) \{ (\pbar+as)^2 +a^2 \mes \} } - 
( s=0\,\,\,part)
\nonumber \\
&=&\frac{1}{4} y^2 \{s^2 +6\mes   \}\cuv.  
\label{eqn:logsh}
\ee 
The four particle contributions of 
$\svpd  \gm V \gm V \svm$ 
(the first term of Eq. (\ref{eqn:gmvgmvb}).) leads to the same 
logarithmic divergence as Eq. (\ref{eqn:logsh}) while the logarithmic 
divergence of 
$ \svpd V \gp \gm V  \svm $ is the twice of Eq. (\ref{eqn:logsh}). 
The other terms do not yield any logarithmic divergence. 
Summing up all the contributions we obtain the final expression,  
\be
\Sigma_{HH}(s) = \frac{\sigma_\Phi}{a^2}
   + y^2 \{s^2   +6\mes   \}\cuv  + (\mu,s,M_H \,\,\,dependent\,\,\, term),
\ee
where $\mu$ is the renormalization point and ${\sigma_\Phi}/{a^2}$ is the 
quadratic divergence. 


As a final example of the Yukawa interactions, we consider the 
contribution of the Higgs boson to the electron self-energy:  
\be
\frac{\int {\cal D}{\cal A}
\langle A+ \vert \Omega(p,q)\vert A- \rangle e^{-S(A)} }
{ \int {\cal D}{\cal A}    \langle A+ \vert A- \rangle e^{-S(A)} }. 
\ee
Here the interaction $V$ is given by (\ref{eqn:yukawa}) 
with (\ref{eqn:yukawah}).
Using Eqs. (\ref{eqn:vgpvgpom1}) and 
(\ref{eqn:vgpvgpom2}), we get 
\be 
\svpd V \gp  V \gp  \nopq \svm = -\smp \sigpq \splusq,  
\ee
where 
\be
\sigpq &=& -\tpidelta{p-q}\frac{1}{2}y^2 \int_{k} D(k-p)
\fwppke^2 
\left( 
\begin{array}{cc}
T_+(k,e)_{RR} &  T_+(k,e)_{RL} \\
T_+(k,e)_{LR} &  T_+(k,e)_{LL}
\end{array}
\right)
\nonumber \\
&=& 
\tpidelta{p-q}\frac{a}{4\lambda}
\left( 
\begin{array}{cc}
\Sigma_+(p,e)_{LL} &  \Sigma_+(p,e)_{LR} \\
\Sigma_+(p,e)_{RL} &  \Sigma_+(p,e)_{RR}
\end{array}
\right). 
\ee
Here each $\Sigma_+(p,e)$ is evaluated as before by rescaling the loop 
momentum and we obtain  
\be 
\Sigma_+(p,e)_{LL (RR) } &=& 
\frac{1}{a}\{ \sigma_1 + \gfive \sigma_2\} 
+ 
P_{R (L)}\frac{1}{4} y^2 i \notp \cuv 
\nonumber \\
&+& (p,m_e \,\,\, and \,\,\,\mu,\,\,\,dependent\,\,\,term), 
\nonumber \\
\Sigma_+(p,e)_{LR (RL) } &=& - P_{R (L)}\frac{1}{2}y^2 \me \cuv.   
\label{eqn:sfh}
\ee
The other contributions are evaluated in the same way. 
The four particle contribution (the first term of 
(\ref{eqn:omgmvgmv2})) and $\sigpq_\pm$ in Eq. (\ref{eqn:vgpomgmv}) 
give rise to the logarithmic divergences same to Eq. (\ref{eqn:sfh}), 
while the other terms do not lead any logarithmic divergences.  
The expression (\ref{eqn:sfh}) 
shows that the chiral symmetry is preserved after taking into account  
the Yukawa interactions and the Dirac mass is 
renormalized multiplicatively.

\setcounter{equation}{0}

\section{Gauge Interaction}

\indent

Next we consider the gauge interactions. 
First we consider the contribution of the electron to the 
vacuum polarization \cite{over,vac1,vac2} of the $Z$ boson:   
\be 
\frac{\int {\cal D}{\cal A} {\cal D} \Phi 
\zmu(s)\znu(t) \langle A + \vert  A- \rangle e^{-S(A)} }
{ \int {\cal D}{\cal A}    \langle A+ \vert A- \rangle e^{-S(A)} },
\ee
where $\langle A + \vert  A- \rangle$ 
should be expanded to the second order in the gauge couplings: 
\be 
\svapd A-\rangle &=& 
\Bigr\{ 1  -  \frac{1}{2} \svpd V \gp^2  V \svp  
- \frac{1}{2} \svmd V \gm^2  V \svm \Bigl\} \svpd -\rangle
+\svpd V \gp  \svm +
\svpd \gm V \svm 
\nonumber\\ 
&+&
\svpd V \gp \gm V  \svm +
\svpd V \gp  V \gp   \svm +
\svpd  \gm V \gm V   \svm. 
\label{eqn:vacpol}
\ee
First we consider the tadpole type corrections 
$\svpd V \gp  \svm$ and $\svpd \gm V \svm $. In this case the 
interaction $V$ is given by 
\be
V=\frac{1}{2}a g^2_Z \int_{p,s,q} \psibar(p)\sum_{\mu\nu}
V_{2\mu}(p+q)
\left(
\begin{array}{ccc}
c_n & 0   & 0    \\
0   & c_L & 0    \\
0   & 0   & c_R
\end{array}
\right)
\zmu(s)\znu(p-s-q)\psi(q).
\ee
Using the formulae (\ref{eqn:vgpb}) and (\ref{eqn:gmvb}), 
the contribution of the electron is given by    
\be 
& &\svpd V \gp  \svm =\frac{1}{2}\sum_{\mu\nu} \int_{s}
\anu(s) \amu(s) \Pi_{\mu\nu}(0),
\nonumber \\
& &\Pi_{\mu\nu}(0)=-\delmunu g^2_Z a \int_{p} 
\frac{1}{2\omp} Tr V_{2\mu} (2p) 
\{ c^2_L T_{+}(p,e)_{LL} + c^2_R T_{+}(p,e)_{RR} \}.
\label{eqn:vactad}
\ee
After rescaling the loop momentum $p \rightarrow \bar{p} = ap$, 
we find that $ \Pi_{\mu\nu}(0)$ is quadratically divergent and is not 
gauge invariant. 
This type of contribution should be cancelled when all the 
terms in Eq. (\ref{eqn:vacpol}) are taken into account. 
At the one-loop level, this is guaranteed by the fact that the real part of 
the overlap is gauge invariant \cite{over,vac1,vac2}. 
Here we demonstrate this cancellation in the limit of the Wilson parameter 
$r \rightarrow 0$ \cite{kawai}.  
In this limit,
\be
\svpd V \gp  \svm +
\svpd \gm V \svm 
\rightarrow \Pi_{\mu\nu}(0)=
- \delmunu \frac{\bar{g}^2}{a^2}  
\int_{\pbar} \ptildebar^2_\mu 
\Bigl\{
\fracp -\fracom 
\Bigr\},
\label{eqn:vactad2}
\ee
with $\bar{g}^2= (c^2_L + c^2_R) g^2_Z$. 
The other contributions are,
\be
& & - \frac{1}{2} \svpd V \gp^2  V \svp,\,\,\,
-\frac{1}{2} \svmd V \gm^2  V \svm
\rightarrow
\Pi_{\mu\nu}(0) 
=
-  \delmunu \frac{\bar{g}^2}{4a^2}  
\int_{p} c^2_\mu 
\Bigl\{
\fracom
- \frac{1}{\tilde{\omega}^4}\ptildebar^2_\mu 
\Bigr\},
\nonumber \\
& &\svpd V \gp  V \gp   \svm,\,\,\, \svpd  \gm V \gm V   \svm
\nonumber \\
& &\,\,\,\,\,\rightarrow 
\Pi_{\mu\nu}(0)
= 
\delmunu  \frac{\bar{g}^2}{4a^2}  
\int_{p} c^2_\mu 
\Bigl\{
\fracp
-\fracom
-\frac{2}{\ptildebar^4}\ptilde^2_\mu 
+\frac{3}{\tilde{\omega}^4}\ptilde^2_\mu 
-\frac{1}{\tilde{\omega}^2\ptildebar^2}\ptildebar^2_\mu 
\Bigr\},
\nonumber \\
& &\svpd V \gp \gm V  \svm 
\rightarrow
\Pi_{\mu\nu}(0)=
\delmunu \frac{\bar{g}^2}{2a^2}  
\int_{p} c^2_\mu 
\Bigl\{
\fracp
-\frac{2}{\ptildebar^4}\ptildebar^2_\mu 
+\frac{1}{\tilde{\omega}^2\ptildebar^2}\ptildebar^2_\mu 
\Bigr\},
\label{eqn:vacself}
\ee
where $c_\mu =\cos(\pbar_\mu)$. Summing up the contributions 
(\ref{eqn:vacself}) the terms proportional to 
$1/(\tilde{\omega}^2\ptildebar^2)$ are canceled with each other. 
For the terms proportional to $c_\mu\ptildebar_\mu / {\tilde{\omega}^4}$, 
using the relation
\be 
\frac{\partial}{\partial \pbar_\mu} \fracom 
=-\frac{2 c_\mu\ptildebar_\mu}{\tilde{\omega}^4}
\ee
and integrating by part, it becomes the second term of Eq. (\ref{eqn:vactad2}) 
with the opposite sign. In this way, the gauge non-invariant 
quadratic divergences cancel with each other in the vacuum 
polarization. The logarithmically divergent part of $\Pi_{\mu\nu}(s)$ 
is evaluated in a similar way to $\Sigma_{HH}(s)$, and the result is 
\be
\Pi_{\mu\nu}(s)=g^2_Z \Bigl[ 
(c^2_L + c^2_R)
\Bigl\{  
\frac{2}{3}(s^2\delmunu -s_\mu s_\nu) + 2\mes \delmunu
\Bigr\} 
-4  c_L c_R \mes \delmunu
\Bigr] \cuv,
\ee
which also satisfies the transverse condition up the spontaneous 
symmetry breaking factor. 

Next we consider the contribution of the $Z$ boson to the 
electron self-energy. 
The four particle contribution in $\svpd V \gp  V \gp  \nopq \svm$ 
is given by   
\be
\sigpq&=&-\tpidelta{p-q}\sum_{\mu\nu}e^{-i(p-q)a/2} g^2_Z
\int_{k} \fwppke^2 
V_{1\mu}(p+k)
\nonumber \\
&\times &\left(
\begin{array}{cc}
c^2_L  T_+(p,e_V)_{LL} & c_Lc_R T_+(p,e_V)_{LR} \\
c_Lc_R T_+(p,e_V)_{RL} & c^2_R T_+(p,e_V)_{RR}
\end{array}
\right)
V_{1\nu}(k+p)D_{\mu\nu}(k-p)
\\
&=&\tpidelta{p-q}\frac{a}{4\lambda} 
\left(
\begin{array}{cc}
c^2_L  \Sigma_{LL}(p) & c_Lc_R \Sigma_{LR}(p) \\
c_Lc_R \Sigma_{RL}(p) & c^2_R  \Sigma_{RR}(p)
\end{array}
\right),
\label{eqn:sfg}
\ee
where 
\be 
& &\Sigma_{LL(RR)}(p) = \frac{1}{a}\{\sigma_1 + \sigma_2 \gfive  \}
+ P_{R(L)} i\notp \cuv,
\nonumber \\
& &\Sigma_{LR(RL)}(p) = P_{R(L)} 4 \me \cuv,
\label{eqn:sfg2} 
\ee
in the t'Hooft-Feynman gauge ($\alpha_Z =1)$.  
The other terms in Eq. (\ref{eqn:loopf}) lead to similar contributions, 
and the logarithmic divergences are obtained only from 
$\sigpq_\pm$ in Eq. (\ref{eqn:vgpomgmv}) and 
the first term of Eq. (\ref{eqn:omgmvgmv2}). 
The expression (\ref{eqn:sfg}) with Eq. (\ref{eqn:sfg2}) 
shows that the fermions are correctly renormalized. 

\setcounter{equation}{0}

\section{Discussion}

\indent

We have developed lattice perturbation theory for the 
Yukawa and gauge interactions in the overlap formulation 
taking a realistic example, the Standard Model, and we have analyzed one 
and two point functions for the fermions, scalars and 
gauge bosons at the one loop level.  
We also gave the compact 
formulae for the self-energies, 
from which both the divergent and finite parts can be 
further analyzed at the one loop level.    

Even though we established the lattice perturbation theory 
for realistic theories and demonstrated the analysis up to the one loop level, 
there remain issues still to be clarified and which did not appear in our 
analysis. 
The unclear point for a regularization of a general chiral gauge 
theory is the role of the imaginary part of $\svapd A-\ran$ 
(or equivalently, the phase convention of the vacuum states $\svapm$),
which can be a functional of the gauge fields 
(such an effect did not appear at the one loop level). 
For the anomaly free case, such effects are suppressed by a positive 
power of the lattice spacing in the expression for $\svapd A-\ran$. 
However, as is shown in the calculation of the vacuum polarization, 
the interactions suppressed by a positive power of the lattice spacing 
play important roles in keeping the gauge invariance at loop levels, 
and the imaginary part of $\svapd A-\ran$, even if suppressed by a positive 
power of the lattice spacing, 
may affect the gauge invariance at higher orders. 

For Higgs models and vector-like theories there will be no such 
subtle points and our analysis suggests that this formulation is feasible 
for regularizing Higgs models, as well as vector like gauge theory 
\cite{vector}, with the advantage of respecting the chiral symmetry. 

\section*{Acknowledgment} 

I would like to thank Y.~Kikukawa and 
S.~Randjbar-Daemi for discussions, and G.~Gyku and 
M.~O'Loughlin for comments on the manuscript.


\renewcommand{\theequation}{\Alph{section}.\arabic{equation}}
\setcounter{equation}{0}
\setcounter{section}{1}
\section*{Appendix A.}

\indent

In this appendix, we present the general formulae to compute the 
self-energies of the fermions and bosons (both scalar and vector bosons) 
in the overlap formulation at the one loop level.  
First we list the propagators appearing in the 
expression:
\be
& &\splusp = 
\frac{1}{\cbep}\sum_{\sigma_f}\upp \umbp,  
\nonumber \\
& &
\smp =  
\frac{1}{\cbep}\sum_{\sigma_f}\vmp \vpbp,  
\nonumber \\
& &P_{\pm}(p,f) = \sum_{\sigma_f}\upmp  \upmbp,\,\,\, 
N_{\pm}(p,f) = \sum_{\sigma_f}\vpmp  \vpmbp. 
\nonumber \\
\label{eqn:prospn}
\ee
In the above, $S_\pm$ develop poles near the center of the Brillouin 
zone. The projection operator 
$P_\pm (N_\pm) = \{ {\bf 1_f} \wpmp +(-) H_{\pm}(p,f)\}\gfive/2\wpmp $ 
do not have any poles. (Here ${\bf 1_f}=diag(0,1,1)$ for the electron, 
for example, and $H_{\pm}(p,f)$ is the block diagonalized part of the one 
particle Hamiltonian (\ref{eqn:h}) acting on the flavor $f$.)  
We find it convenient to introduce other two propagators:   
\be
\tplusp &=& 
\frac{\sin\beta(p,f)}{\cbep} \sum_{\sigma_f}
\upp \vpbp  
\nonumber\\
&=& -\splusp +\pplusp = \smp -\nmp,
\nonumber \\
\tmp &=&  
\frac{\sin\beta(p,f)}{\cbep}\sum_{\sigma_f}\vmp \umbp  
\nonumber\\
&=& -\smp +\pmp = \splusp -\nplusp.
\label{eqn:prot}
\ee
Explicit expressions of the $S_\pm$, $P_\pm$ and $N_\pm$ for the Dirac type propagators are, 
\be
S_{\pm}(p,f_V) &=& \frac{1}{Z_\pm}  
\left( 
\begin{array}{cc}
K_\pm (p,f_V)_{LL}   & M_\pm (p,f_V)_{LR}  \\
M_\pm (p,f_V)_{RL} \ & K_\pm (p,f_V)_{LL} 
\end{array}
\right), 
\label{eqn:spmex}
\ee
where
\be
Z_\pm 
&=& (\omp \pm \xp)(\omm \pm \xm) + \ptilde^2 +\mfs, 
\nonumber \\
K_+(p,f_V)_{LL(RR)} 
&=&
P_{L(R)} (\omp +\xp) 
\Bigl\{   
-i \fnot{\ptilde} + T_{f_{L(R)}}(\omm + \xm) 
\Bigr\}
\nonumber \\
&+& P_{R(L)} (\omm +\xm) 
\Bigl\{   
-i \fnot{\ptilde} - T_{f_{L(R)}}(\omm - \xm) 
\Bigr\},
\nonumber \\
M_+(p,f_V)_{LR(RL)} &=&
P_{L(R)} \mf (\omp +\xp) 
+ P_{R(L)} \mf (\omm +\xm), 
\label{eqn:spmex2}
\ee
\be
& &P_{\pm}(p,f)= \frac{1}{2\wpmp} 
\left( 
\begin{array}{cc}
\wpmp\gfive  - i \fnot{\ptilde} + T_{f_L} \xpm   & m_f \\
 m_f & \wpmp\gfive - i \fnot{\ptilde} + T_{f_R} \xpm 
\end{array}
\right), 
\nonumber \\ 
& &N_{\pm}(p,f)= \frac{1}{2\wpmp} 
\left( 
\begin{array}{cc}
\wpmp \gfive  + i \fnot{\ptilde} - T_{f_L} \xpm   & - m_f \\
- m_f & \wpmp \gfive  + i \fnot{\ptilde} - T_{f_R} \xpm \}
\end{array}
\right). 
\nonumber \\
\label{eqn:pnex}
\ee

Now we show the expressions of $\svapd \nopq \svam$ and $\svapd A-\rangle$ 
which are necessary to compute the fermion and boson self-energies. 
To compute these factors, we only consider the one-particle irreducible 
contributions, and we use the interaction vertex 
$\gampq$ (which is of the matrix form) defined by, 
\be
V=\int_{p,q} \psibar(p) \gampq \psi(q).    
\ee

For the calculation up to the one loop level, 
$\svapd \nopq \svam$ is expanded up to the second order in $V$: 
\be
\svapd \nopq \svam &=& 
\svpd V \gp \nopq \svm + 
\svpd \nopq \gm V \svm 
+
\svpd V \gp \nopq \gm V  \svm 
\nonumber \\
&+&
\svpd V \gp  V \gp  \nopq \svm +
\svpd \nopq \gm V \gm V   \svm, 
\label{eqn:loopf}
\ee 
where the first two terms are evaluated as, 
\be 
& &\svpd V \gp \nopq \svm = \smp \fwppq \gampq \splusq, 
\nonumber \\
& &\svpd \nopq \gm V \svm = \splusp \fwmpq \gampq \smq. 
\label{eqn:tadf}
\ee
The third term is, 
\be 
& & \svpd V \gp \nopq \gm V  \svm =
\splusp \sigpq_+ \splusq + \smp \sigpq_- \smq,
\nonumber \\ 
& &\sigpq_+ = \sumk \fwmpk \fwpkq \gampk \{-\smk \} \gamkq, 
\nonumber \\
& &\sigpq_- = \sumk \fwppk \fwmkq \gampk \splusk \gamkq. 
\nonumber \\
\label{eqn:vgpomgmv}
\ee
To evaluate these terms, only the two particle states contribute 
in the sum of the excited states of $G_\pm$. 
To evaluate the last two terms in Eq. (\ref{eqn:loopf}), both two and four 
particle states have to be taken into account in $G_\pm$. The results are, 
\be
& &\svpd V \gp  V \gp  \nopq \svm = -\smp \sigpq \splusq, 
\label{eqn:vgpvgpom1}
\\
& &\sigpq= -\sumk \fwppk \fwpkq \gampk \tplusk \gamkq 
\nonumber \\
& &- \fwppq \sumk \gampk \Bigr\{ 
- \frac{\pplusk}{\wppk}  + \frac{\nplusk}{\wpkq} \Bigl\}\gamkq.
\nonumber \\
\label{eqn:vgpvgpom2}
\ee
\be
& &\svpd \nopq \gm V \gm V   \svm = \splusp \sigpq \{- \smq \},
\\
& &\sigpq= \sumk \fwmpk \fwmkq \gampk \tmk  \gamkq 
\nonumber\\
& &-
\fwmpq \sumk \gampk \Bigr\{
- \frac{\pmk}{\wmpk}  + \frac{\nmk}{\wmkq} \Bigl\} \gamkq.
\nonumber \\
\label{eqn:omgmvgmv2}
\ee
In the above expressions the first terms of $\Sigma(p,q)$ are the 
contributions of the four 
particle states, while the second parts are those of the two 
particle states. 

To compute the boson self-energies in the one-loop, 
$\svapd A-\rangle$ should be expanded up to the second order in $V$: 
\be
\svapd A-\rangle &=& 
\Bigr\{ 1  -  \frac{1}{2} \svpd V \gp^2  V \svp  
- \frac{1}{2} \svmd V \gm^2  V \svm \Bigl\} \svpd -\rangle
+\svpd V \gp  \svm +
\svpd \gm V \svm 
\nonumber \\
& &
+\svpd V \gp \gm V  \svm +
\svpd V \gp  V \gp   \svm +
\svpd  \gm V \gm V   \svm,
\label{eqn:loopb}
\ee
where the first term proportional to $\svpd -\rangle$ comes from the 
normalization factors $\alpha_\pm(A)$. Each term gives rise to the 
following expression: 
\be  
& &\svpd V \gp  \svm = \sumk (-1) \frac{1}{2 \wpk} 
Tr \Gamma(k,k) \tplusk, 
\\
& &
\svpd \gm V   \svm =
 \sumk \frac{1}{2 \wmk}
Tr \Gamma(k,k) \tmk,
\label{eqn:vgpb}
\ee    
\be
\svpd V \gp^2  V \svm = \sumpq \fwppq^2 
Tr \pplusp \gampq \nplusq\gamqp,  
\label{eqn:gmvb}
\ee
\be
\svmd V \gm^2  V \svm = 
\sumpq \fwmpq^2
Tr \pmp \gampq \nmq\gamqp,
\ee
\be
\svpd V \gp \gm V  \svm &=&  \sumpq \fwppq \fwmpq 
\nonumber\\
& \times& Tr \splusp \gampq \smp \gamqp, 
\ee
\be 
& &
\svpd V \gp  V \gp   \svm = 
-\frac{1}{2} \sumpq \fwppq^2 
Tr \tplusp \gampq \tplusq \gamqp 
\nonumber\\
& &
+\sumpq \frac{1}{2\wpp}  \fwppq 
Tr \tplusp \gampq \{\pplusq - \nplusq   \} \gamqp,  
\nonumber \\
\label{eqn:vgpvgpb}
\ee
\be
& &\svpd  \gm V \gm V   \svm = 
-\frac{1}{2} \sumpq \fwmpq^2 
Tr \tmp \gampq \tmq \gamqp 
\nonumber \\
& &
+\sumpq \frac{1}{2\wpp}  \fwmpq 
Tr \tmp \gampq \{\nmq - \pmq   \} \gamqp. 
\nonumber \\ 
\label{eqn:gmvgmvb}
\ee
In the above expressions, the trace is over the flavor space and the 
Dirac matrices. 
In Eqs. (\ref{eqn:vgpvgpb}) and (\ref{eqn:gmvgmvb}), 
the first terms are the contribution of the four particle states in $G_\pm$, 
while the second terms are those of the two particle states.  
The contributions of the other terms are only from the two particle 
states. 


\setcounter{equation}{0}

\setcounter{section}{2}

\section*{Appendix. B. Gauge and Higgs sector}

\indent

In this appendix we briefly give the lattice formulation  of the 
$SU(2) \times U(1)$ Higgs doublet. Further discussions are found e.g., 
in Refs. \cite{Shrock,montvay1,montvay2}.  
The action of the Higgs doublets  
\be
\Phi(x) = \left(  
\begin{array}{c}
\phi_+(x) \\
\phi_0(x)
\end{array}
\right)
=
\left(  
\begin{array}{c}
i \pi_+(x) \\
\{ v + H(x) -i \pi_3(x) \}/{\sqrt{2}}
\end{array}
\right)
\ee
is given by
\be 
S_E = \int d^4x \Bigr[ 
\phid(x) ( - \Box + \mu^2 ) \Phi (x) + 
\lamphi \{ \phid(x)  \Phi(x) \}^2  \Bigl].
\ee
The gauge invariant lattice action is obtained by discretising it and 
inserting the link variable:
\be
S_E = a^4 \sum_{n} 
\Bigr[
-\frac{1}{a^2}\sum_{\mu} 
2Re \{ \phid(n) W_{n,n+\muhat}\Phi(n+\muhat) \} 
+ \frac{1}{a^2} 2d \phid(n)\Phi(n)
+\lamphi \{ \phid(n)  \Phi(n)    \}^2 
\Bigl].
\nonumber \\
\label{eqn:shlat}
\ee
The lattice action of the gauge fields is given by 
\be
S_G = 
\frac{2N}{g^2}\sum_{P} 
\Bigl\{
1- \frac{1}{2N} Tr\Bigl( U_P + U^\dagger_P \Bigr) 
\Bigr\}
+
\frac{1}{g'^2}\sum_{P} 
\Bigl\{
1- \frac{1}{2} Tr\Bigl( V_P + V^\dagger_P \Bigr) 
\Bigr\},
\ee
where $U_P$ and $V_P$ stand for the products of the link variables 
around a plaquette $P$, the sum is over the plaquettes and 
$N=2$ for $SU(2)$.  
Expanding the link variables in terms of the gauge couplings 
in the action (\ref{eqn:shlat}), we find the mass eigenstates of the 
neutral gauge bosons as  
$\zmun = \cw \wtmun -\sw \bmun$ and $ 
\amun = \sw \wtmun + \cw \bmun$ with $\tan\theta_W = g'/g$. 
For the weak coupling expansion, it is convenient to take the 
covariant gauge with the gauge fixing term, 
\be 
S_{GF} &=& 
\frac{1}{\alphaw} a^4 \sum_{n} 
\Bigr\{ \sum_{\mu} \delmu^L \wpmun + \alphaw \mw \pi_+(n) \Bigl\}^\dagger  
\Bigr\{ \sum_{\mu} \delmu^L \wpmun + \alphaw \mw \pi_+(n) \Bigl\} 
\nonumber \\
&+&
\frac{1}{2\alphaz} a^4 \sum_{n} 
\Bigr\{ \sum_{\mu} \delmu^L \zmun + \alphaz \mz \pi_3(n) \Bigl\}^2
+
\frac{1}{2\alphaa} a^4 \sum_{n} 
\Bigr\{ \sum_{\mu} \delmu^L \amun \Bigl\}^2, 
\ee
where $\partial^L_\mu $ and $\partial^R_\mu $ are the left and right 
lattice derivatives. 
Taking into account the lattice action of the Higgs fields, gauge fields and 
the gauge fixing terms, we obtain following free parts,   
\be 
S &=&  a^4 \sum_{n} 
\Bigr\{ 
(\mu^2 + \lamphi v^2 ) \Bigr\{ v H(n) + \frac{1}{2} H^2(n)+    
\frac{1}{2} \pi^2_3(n)   + \pi_+(n) \pi_-(n) \Bigl\} 
\nonumber\\ 
&+&
\frac{1}{2} H(n) (- \sum_{\rho}\partial^L_\rho \partial^R_\rho  + \mhs ) H(n)
+ \frac{1}{2} \pi_3(n) (- \sum_{\rho}\partial^L_\rho \partial^R_\rho + \alphaz\mzs ) \pi_3(n) 
\nonumber \\
&+& 
\pi_-(n) (- \sum_{\rho}\partial^L_\rho \partial^R_\rho + \alphaw\mws ) 
\pi_+(n) 
+ \frac{1}{2} \zmun G_{\mu\nu}(\alphaz,\mz)  \zmun
\nonumber \\
&+&
\frac{1}{2} \amun G_{\mu\nu}(\alphaa,0) \amun
+\wmmun G_{\mu\nu}(\alphaw,\mw)  \wpmun
\Bigl\},
\label{eqn:shg}
\ee
where   
\be
G_{\mu\nu}(\alpha,M) = 
\delmunu ( - \sum_{\rho}\partial^L_\rho \partial^R_\rho  + M^2) 
+(1- \frac{1}{\alpha} )\delmu^R \delnu^L. 
\ee
The path integral measure and the Faddeev-Popov ghost terms 
are not necessary in our analysis and do not present here. 
The propagators are computed from the action (\ref{eqn:shg}).    
Performing the Fourier transformations 
(see, Eq. (\ref{eqn:fourier})), 
we obtain the propagator for 
scalars as 
\be 
\lan \phi(p) \phi^\dagger(q) \ran = 
\tpidelta{p+q}  D(p,M),\,\,\,D(p,M)=\frac{1}{\phat^2+ M^2}.
\ee 
For the gauge fields, 
\be 
& &\lan \amu(p)\anu(q) \ran =\tpidelta{p+q} D_{\mu\nu}(p,0)   
e^{-i{(p+q)_\nu}a/2},
\nonumber \\ 
& &\lan \zmu(p)\znu(q) \ran =\tpidelta{p+q} D_{\mu\nu}(p,\mz)
e^{-i{(p+q)_\nu}a/2},
\nonumber\\
& & 
\lan \wpmu(p)\wmnu(q) \ran =\tpidelta{p+q} D_{\mu\nu}(p,\mw)
e^{-i{(p+q)_\nu}a/2},
\nonumber \\
& &
D_{\mu\nu}(p,M)=D(p,M) 
\Bigl\{ 
\delmunu - (1-\alpha) \frac{\phat_\mu\phat_\nu}{\phat^2+\alpha M^2}       
\Bigr\}.   
\ee


\begin{thebibliography}{99}
\bibitem{phase}
Nontrivial phases of 
$\alpha_{\pm}(A) $ appear at the order ${\cal O}(g^3)$ \cite{kikukawa,n} 
and do not affect the calculations of the propagator at one loop level.
\end{thebibliography}
\end{document}